\documentclass[showpacs,prb,superscriptaddress,twocolumn]{revtex4}
\usepackage{textcomp}
\usepackage{amsmath}
\usepackage{amssymb}
\usepackage{graphicx} 

\begin{document}

\title{Electronic implementations of Interaction-Free Measurements}

\author{L. Chirolli}
\email{luca.chirolli@uni-konstanz.de}
\affiliation{Department of Physics, University of Konstanz, 
D-78457 Konstanz, Germany}

\author{E. Strambini}
\affiliation{NEST, Scuola Normale Superiore and Istituto Nanoscienze -- CNR, 
		Pisa, Italy}

\author{V. Giovannetti}
\affiliation{NEST, Scuola Normale Superiore and Istituto Nanoscienze -- CNR, 
		Pisa, Italy}
		
\author{F. Taddei}
\affiliation{NEST, Scuola Normale Superiore and Istituto Nanoscienze -- CNR, 
		Pisa,  Italy}

\author{V. Piazza}
\affiliation{NEST, Scuola Normale Superiore and Istituto Nanoscienze -- CNR, 
		Pisa,  Italy}

\author{R. Fazio}
\affiliation{NEST, Scuola Normale Superiore and Istituto Nanoscienze -- CNR, 
		Pisa,  Italy}

\author{F. Beltram}
\affiliation{NEST, Scuola Normale Superiore and Istituto Nanoscienze -- CNR, 
		Pisa, Italy}

\author{G. Burkard}
\affiliation{Department of Physics, University of Konstanz, 
D-78457 Konstanz, Germany}

\begin{abstract}
Three different implementations of interaction-free measurements (IFMs) in solid-state nanodevices are discussed. The first one is based on a series of concatenated Mach-Zehnder interferometers, in analogy to optical-IFM setups. The second one consists of a single interferometer and concatenation is achieved in the time domain making use of a quantized electron emitter.
The third implementation consists of an asymmetric Aharonov-Bohm ring.
For all three cases we show that the presence of a dephasing source acting on one arm of the interferometer can be detected without degrading the coherence of the measured current.
Electronic implementations of IFMs in nanoelectronics may play a fundamental role as very accurate and noninvasive measuring schemes for quantum devices. 
\end{abstract}


\maketitle

\section{Introduction}

{\em Interaction-free measurements} 
(IFMs) were first introduced by ÊElitzur and Vaidman \cite{ElitzurVaidman}, who showed 
that the laws of quantum mechanics allow to reveal the presence of an object without disturbing it. 
The original proposal exploited the coherent splitting and the subsequent recombination of the Ê
wave-function of a photon entering a ÊMach-Zehnder (MZ) interferometer. The disturbance induced by the 
object placed in one of the two arms of the interferometer (an absorber in the original proposal)
manifests itself in the properties of the outgoing photon flux. Upon suitable setting of the interferometer parameters
it was shown that even without absorption taking placeÊ its mere possibility does
modify the state of the particle emerging from the interferometer. As a result an external observer will be able 
to gather information about the presence or absence of the absorber, without Êthe photon being actually absorbed. 
The maximal success probability was bound to be 50\% in the original proposal. A way to improve the 
efficiency of the scheme was put forward by P. Kwiat {\it et al.} \cite{KwiatPRL95}, who suggested to use 
coherently  repeated interrogations. In their scheme a photon was repeatedly Êsent into a MZ Êinterferometer, 
with an absorber Êplaced in one of the two arms. By properly tuning the MZ Êphase it was shown that it is possible to enhance the efficiency of the setup arbitrarily Êclose to 1. Such a scheme can be thought as an application of Êa discrete form of the quantum Zeno effect\cite{ZENO} since every step can be considered as a measurement accompanied by state reduction. 

IFMs Êwere experimentally 
realized using single-photon sources~\cite{KwiatPRL95,TsegayePRA98, WhitePRA98,vanVoorthuysen96} and in neutron
interferometry \cite{HafnerPhysLettA97}.
The enhanced efficiency of concatenated MZ interferometers schemes
was tested in Ref.~[\onlinecite{KwiatPRL99}] with a demonstrated improvement up to Ê73\%. Its application was extended to the case of
semitransparent objects with classical light~\cite{JangPRA99,PAVICIC,INOUE,GiovannettiOptExp06}. 
An important consequence Êof these works is that IFM can be interpreted in terms of deterioration of a resonance condition~\cite{PAVICIC} which does not necessarily need a quantum description (``classical" optical coherence is sufficient), at least for these optical realizations. 

The implementation of IFM in electronic devices deserves in our opinion a careful scrutiny since it constitutes an ideal 
test-bed for the study of quantum-control and quantum-mechanics phenomena in mesoscopic systems.
It is worth noting that, Êdifferently from the optical case, for electronic systems there is no corresponding classical model to realize an IFM. 
In recent years advances in device fabrication opened the way to the observation of interference phenomena in electronic-transport Êexperiments, suggesting important opportunities for a variety of applications.
The achievements obtained in the context Êof Êtwo-dimensional electron gases in the integer quantum Hall effect regime~\cite{GOERBIG} are of particular interest for what follows. Here, various experimental realizations of the MZ \cite{JiNature03,NederPRL06,LitvinPRB07, RoulleauPRB07,NederNatPh07} and Hanbury-Brown-Twiss interferometers \cite{SamuelssonPRL04,NederNatur07} were successfully implemented.
In addition, quantized electron emitters were recently realized~\cite{Moskalets08, Splettstoesser08,Gabelli06, Feve07}. 
The possibility to extend ÊIFM Êto Êelectronic systems seems therefore now at reach, paving the way to the development ofÊ novel 
non-invasive measurement schemes in mesoscopic systems, with possible important implications for quantum information processing. 

A first application of IFM strategies to electronic systems Êwas proposed in Ref. [\onlinecite{PARA}] to detect the presence of a current 
pulse in a circuit by monitoring Êthe state of a Êsuperconducting qubit coupled to the circuit, Êwithout any Ê
energy exchange between the two. ÊSubsequently, Êin the very same spirit of the original works~\cite{ElitzurVaidman,KwiatPRL95},
it was Êshown how to employ IFM to detect with unitary efficiency a source of noise acting on one arm of an Aharonov-Bohm (AB) Ê
chiral ring without affecting the transmitted and reflected currents~\cite{NOI09}. ÊIn view of its (unavoidable) presence in nanoelectronics, 
the proposal focused on the Êdetection of Êexternal random fluctuating electric or magnetic fields, which represents the most common source of 
noise in nanoscale quantum devices~\cite{MarquardtPRB02,MarquardtPRB03,MarquardtPRL04}. 
Therefore, in Ref.~[\onlinecite{NOI09}] 
a classical fluctuating electrical field that randomizes the phase of the electron traveling through it played the role of the absorber in optical schemes~\cite{ElitzurVaidman,KwiatPRL95,TsegayePRA98,WhitePRA98,vanVoorthuysen96,HafnerPhysLettA97,KwiatPRL99,JangPRA99,
PAVICIC,INOUE,GiovannettiOptExp06}. The resulting apparatus operates as a sort of Ê{\em quantum fuse} which opens or closes a contact depending on the presence or on the 
absence of the dephasing source. The results presented in Ref.~[\onlinecite{NOI09}] Êshow that the mechanism underlying 
the IFM does not depend, to a large extent, on the type of disturbance which is induced in the interferometer. 

In the present paper we extend our previous work \cite{NOI09} on the electronic version of the IFM in several ways. ÊFirst of all we introduce two alternative IFM implementations 
based on the integer quantum Hall effect. ÊThe first scheme closely resembles the optical setup of Ref.~[\onlinecite{KwiatPRL99}] and 
uses a Êrecent proposal~\cite{GiovannettiPRB08} for realizing Êconcatenated MZ interferometers. Ê
The second scheme Êinstead is based on the standard quantum Hall interferometric architecture~\cite{JiNature03,NederPRL06,LitvinPRB07,
RoulleauPRB07,NederNatPh07,SamuelssonPRL04,NederNatur07} and assumes the presence of a quantized electron 
emitter~\cite{Gabelli06, Feve07,Moskalets08, Splettstoesser08}. ÊAs in Ref.~[\onlinecite{NOI09}], both setups Êare shown to be capable of 
detecting the presence of a localized Êdephasing source without affecting Êthe coherence of the probing signals. 
Finally we review the AB-ring Êimplementation of ÊRef.~[\onlinecite{NOI09}] and provide a detailed characterization of the scheme.

The paper is organized as follows. In Sec.~\ref{Sec2} we present a noise-sensitive coherent electron detector, based on the concatenation 
of several MZ interferometers. We show that we can detect the presence of a dephasing source affecting propagation in one of the interfering electronic paths by measuring the output currents. We then study the coherence of the outgoing signal by computing the fraction of coherent signal and show that an IFM measurement of the dephasing source is achievable. In Sec.~\ref{Sec3} we Êembed the device described in Sec.~\ref{Sec2} 
in a larger Mach-Zehnder interferometer and study the visibility of the output currents, showing how the coherence of the outgoing signal can be experimentally addressed.
In Sec.~\ref{Sec4} we propose an implementation of IFM based on a single Mach-Zehnder interferometer that makes use of a quantized electron source and concatenation in the time domain.
In Sec.~\ref{Sec5} Êwe present a double ring structure in which a small chiral AB ring is embedded in one arm of a larger AB ring.
We show that the current which flows through the whole device is a measure of the coherent character of the detection.

\section{Coherent detection of noise with IFMs}
\label{Sec2}

\begin{figure}[t]
\begin{center}
Ê\includegraphics[width=8cm]{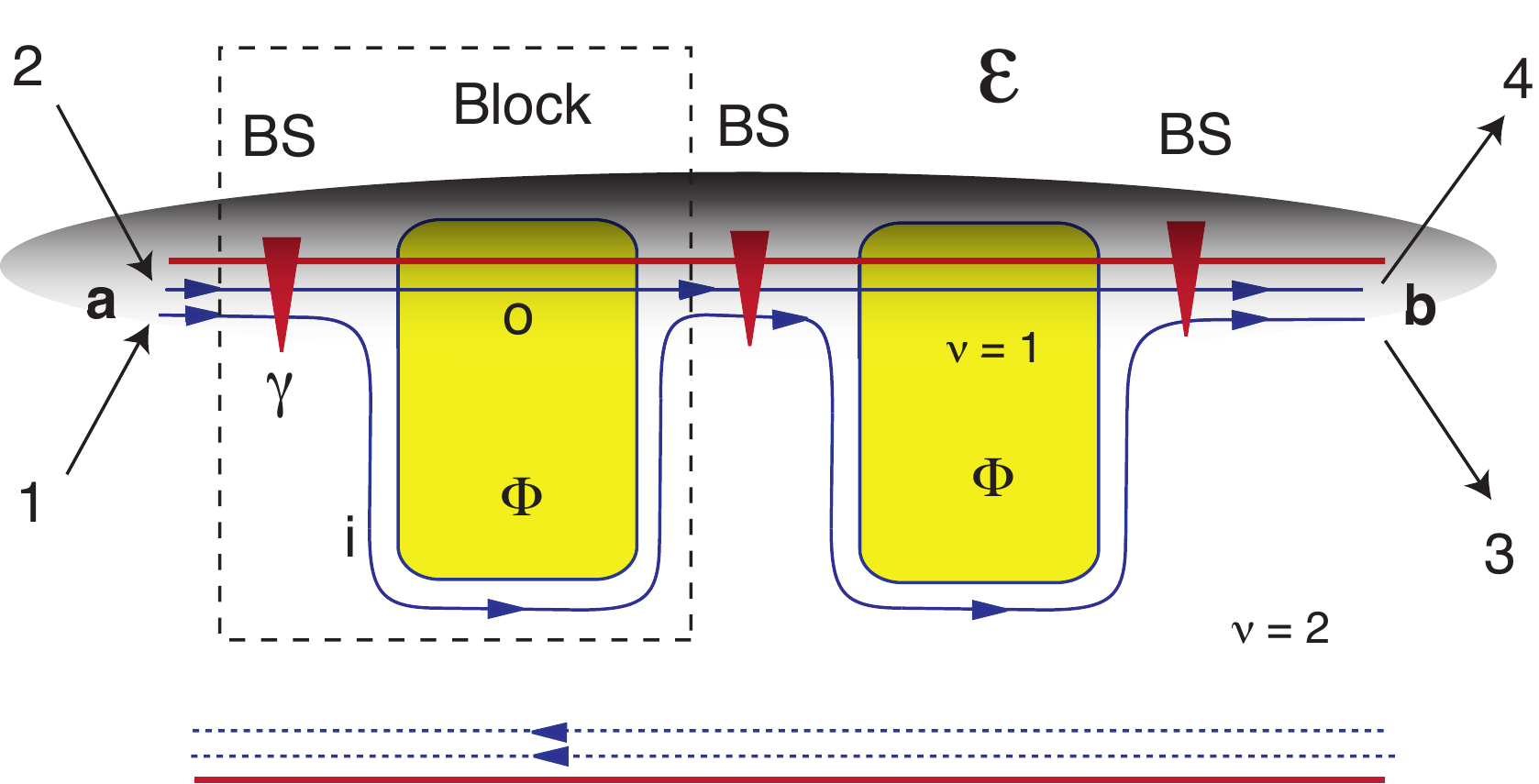}
ÊÊÊ\caption{Schematic illustration of a noise-senitive coherent electron channel 
ÊÊÊ		consisting of $N=2$ representative Êblocks, implemented 					
		in a quantum Hall bar at integer filling $\nu=2$. Incoming electrons in 		
		contact 1 and 2	are represented by their annihilation operators 
		${\bf a}$ and outgoing electrons in contact 3 and 4 by their 				
		annihilation operators ${\bf b}$. ÊEach block is constituted by a beam 		
		splitter (BS) and a phase shifter (PS). Each BS is characterized by a 			
		degree of admixture $\gamma$ and 
		mixes the 	incoming electron in the $i$ and $o$ edge states. The PS 			
		is constituted by an applied top gate (yellow area with filling factor 
		$\nu=1$) that spatially separates the edge channels and introduces 
		a phase difference 
		$\phi$. An external fluctuating field of strength $\epsilon$ (shaded 			
		area) introduces dephasing by randomly shifting the phase of 				
		the electron traveling in the $o$ edge state. \label{Fig1}}
\end{center}
\end{figure}

A straightforward implementation of IFM Êalong the lines developed originally in optics can be realized exploiting the edge-channel interferometric 
architecture of Ref.~[\onlinecite{GiovannettiPRB08}] based on the integer quantum Hall effect at filling factor $\nu=2$. 
The feature of this architecture which is particularly relevant for our purposes is that it Êallows for successive concatenations of different interferometers.
In this scheme, beam splitters (BSs) are realized by introducing a sharp potential barrier which mixes the two edges.
Populating initially only one channel, at the output of a BS we find electrons in a superposition state.
Additional phase shifters (PSs) can be easily realized by spatially separating the two channels with the used of a top gate that can locally change the filling factor to $\nu=1$: only one channel can traverse the region at $\nu=1$ and the other is guided along its edges.
This is schematically shown 
in Fig.~\ref{Fig1}, where a phase difference $\phi$ is introduced between the channels by changing the path of the incoming channel.

Based on this approach we can build an apparatus Êwhich implements an IFM scheme
Along the lines of Êthe optical setup of Ref.~[\onlinecite{KwiatPRL99}].
The proposed device, illustrated in Fig.~\ref{Fig1}, consists 
in a sequence of $N$ interferometric elements, Êin which output edges emerging from the $n$-th interferometer Êare directly 
fed into the input of the $(n+1)$-th one.
As we shall see, the apparatus allows one to
detect the presence of a fluctuating electromagnetic 
field affecting the upper region of the Hall bar
(depicted as a shaded area in Fig.~\ref{Fig1}), Êwithout any coherence loss of the transmitted currents. This is obtained thanks 
to the action of the top gates of the setup which divert the path of the $i$ (inner) channel 
inside the Hall bar Ê(where the fluctuating field is supposed to be absent), and thanks to Êthe coherent mixing between the Ê$i$ Êchannel Êand the $o$ ÊÊ(outer)
channel induced by the BSs.
If Êno dephasing is present Êin the upper region of the bar, then
the electron coherently propagates towards the 
next step, that is nominally equal to the previous one. By properly tuning the degree of admixture of the channel populations, 
it is possible to gradually transfer the electron from the Ê$i$ channel to the $o$ ÊchannelÊat the end of a chain of $N$ interferometers. 
The situation changes completely Êwhen the dephasing field is present in the shaded region of Fig.~\ref{Fig1}. Ê
Indeed, as a result of a random phase shift, 
the part of the wave-function that propagates in the $o$ channel does not coherently add to the one propagating in the $i$ channel. Consequently the gradual transfer of electronic amplitude from Ê$i$ to Ê$o$ does not take place. At the same time the electron that propagates into the channel not exposed to the fluctuating field preserves its 
coherence. The presence or absence of noise is revealed by the electron emerging from lead 3 or 4, respectively and, as we shall clarify in the following, the setup does preserve the coherence of the emerging electronic signal. 

\subsection{Detection of a dephasing noise source}

\begin{figure}[t]
\begin{center}
Ê\includegraphics[width=8cm]{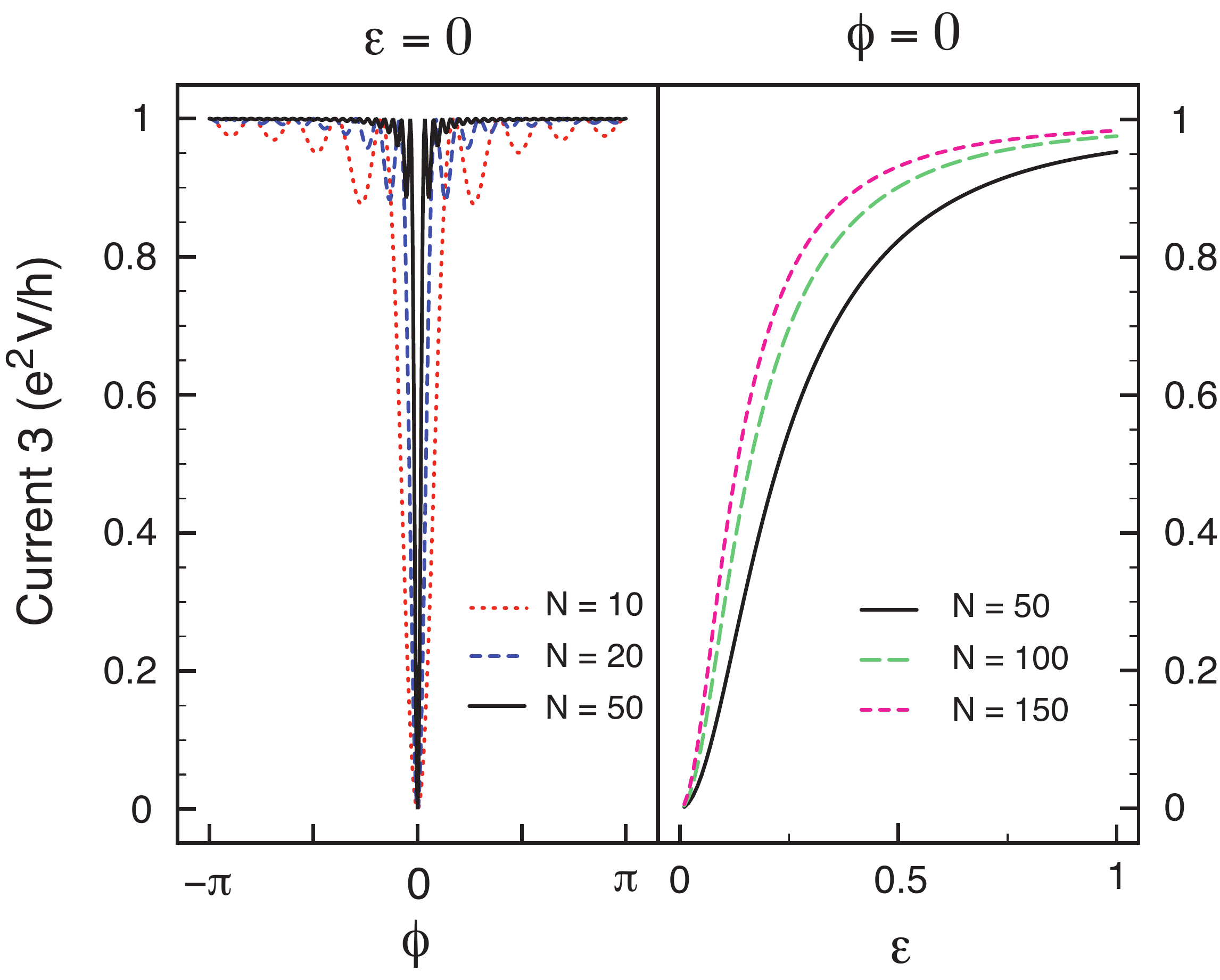}
ÊÊÊ\caption{Current in contact 3 for different number $N$ of blocks. (Left panel) 			
ÊÊÊ		$\langle I_{3,N} \rangle_\delta$ of Eq.~(\ref{eqi3}) versus the phase shift $\phi$ in the coherent case 
ÊÊÊ		$\epsilon=0$. By increasing $N$ a narrow dip arises in the coherent 			
		case for $\phi=0$, and all the current goes out in contact 4. (Right 			
		panel) $\langle I_{3,N} \rangle_\delta$ versus the strength $\epsilon$ of the dephasing field, at 		
		the working point $\phi=0$. As $\epsilon$ increases, the current 			
		tends to go out all from contact 3, thus witnessing the presence of the 		
		dephasing field. \label{Fig2}}
\end{center}
\end{figure}

Electron propagation is described in the Landauer-B\"uttiker formalism of quantum transport~\cite{Buttiker85,Buttiker88}. 
The scattering matrix that describes transport in each block can be written as:
\begin{equation}
S(\delta)=\left(\begin{array}{cc}
e^{i\phi}\cos(\gamma/2) & ie^{i\phi}\sin(\gamma/2)\\
ie^{i\delta}\sin(\gamma/2) & e^{i\delta}\cos(\gamma/2)
\end{array}\right),
\end{equation}
where $0<\gamma<2\pi$ parametrizes the degree of edge-channel mixing introduced by BS, and $\phi$ is the phase shift between the two edge channels.
The presence of a dephasing source is described by a random phase shift $\exp(i\delta)$. 
By using this scattering matrix it is possible to relate electrons exiting the chain of $N$ 
blocks to the incoming ones at the beginning of the chain:
\begin{equation}
{\bf b}=\prod_{i=1}^NS(\delta_i)~{\bf a},
\end{equation}
with ${\bf a}=(a_i,a_o)^T$ being the Fermionic annihilation operator describing 
incoming electrons in leads 1 and 2 (connected to 
channels $i$ and $o$, respectively), and ${\bf b}=(b_i,b_o)^T$ the Fermionic annihilation 
operator describing outgoing electrons (leads 3 and 4). 
Contact 1 is biased at a chemical potential $eV$, reservoirs 
2, 3, and 4 are kept at reference potential. Setting the temperature to zero, Êthe current in contact $3$ is 
\begin{equation}
I_{3,N}=\frac{e^2V}{h}
\left|[S_N]_{11}\right|^2,
\end{equation}
while the current in contact 4 is
\begin{equation}
I_{4,N}=\frac{e^2V}{h}
\left|[S_N]_{21}\right|^2,
\end{equation}
with $S_N=\prod_{i=1}^NS(\delta_i)$. Here we do not take into 
consideration the electron-spin degree of freedom.

The effect of the fluctuating field can be taken into account by averaging the Êphases $\delta_i$ over a generic distribution of width $2\pi\epsilon$ and zero mean. For simplicity we assume a uniform distribution. 
The outgoing currents depend now entirely on the degree of mixing $\gamma$ of edge states in the BS and on the phase shift $\phi$. The average current in contact 3 (4) is given by
\begin{equation}\label{Eq:I3ind}
\langle I_{3(4)}\rangle_{\delta}\equiv\frac{1}{(2\pi\epsilon)^N}
\int_{-\pi\epsilon}^{\pi\epsilon}d\boldsymbol{\delta}~I_{3(4),N},
\end{equation}
with $d\boldsymbol{\delta}=d\delta_1\ldots d\delta_N$. We define the two-component vectors 
${\bf e}_+=(1,0)^T$, ${\bf e}_-=(0,1)^T$, that allow us to express
\begin{eqnarray}
\left|[S_N]_{11}\right|^2&=&{\bf e}_+^TS_N^{\dag}{\bf e}_+{\bf e}_+^TS_N{\bf e}_+,\\
\left|[S_N]_{21}\right|^2&=&{\bf e}_+^TS_N^{\dag}{\bf e}_-{\bf e}_-^TS_N{\bf e}_+.
\end{eqnarray}
Introducing a representation of $2\times2$ matrices in terms of Pauli operators, Ê
concisely written through the Pauli vector $\boldsymbol{\sigma}=(\sigma_0, 
\sigma_1, \sigma_2, \sigma_3)^T$, with $\sigma_0=\openone$, we can write 
${\bf e}_{\pm}{\bf e}_{\pm}^T=(\openone\pm\sigma_Z)/2\equiv{\bf p}_{\pm}\cdot
\boldsymbol{\sigma}$, with $({\bf p}_{\pm})_i={\rm Tr}({\bf e}_{\pm}{\bf e}_{\pm}^T\sigma_i)/2$. 
This allows us Êto calculate the average over phases $\delta_i$ as a matrix product. By defining matrix 
\begin{equation}\label{Eq:DecohMap}
{\cal Q}_{ij}=\frac{1}{2}\int_{-\pi\epsilon}^{\pi\epsilon} \frac{d\delta}{2\pi\epsilon}
{\rm Tr}
\left[S^{\dag}(\delta)\sigma_i S(\delta)\sigma_j\right],
\end{equation}
we can write the zero-temperature average current in output 3 (4) after $N$ blocks as
\begin{equation}
\langle I_{3(4),N}\rangle_{\delta}=\frac{e^2V}{h}
~{\bf p}_{\pm}\cdot{\cal Q}^N
\cdot \left({\bf e}_+^T~\boldsymbol{\sigma}~
{\bf e}_+\right). \label{eqi3}
\end{equation}
We point out that, due to the unitarity of $S(\delta)$, Ê${\cal Q}_{ij}$ defined in Eq.~\ref{Eq:DecohMap} preserves 
the trace. ÊOne can reduce the dimensionality of the problem and work with the Bloch representation of $2\times 2$ density matrices. 

The behavior of the output currents in the limit of large $N$ is obtained by studying the eigenvalues of the $4\times 4$ matrix ${\cal Q}$. Ê
Choosing the working point $\phi=0$, ${\cal Q}$ assumes a diagonal block form that allows a direct solution: 
${\cal Q}=U^{-1}{\rm diag}(1,\sin(\pi\epsilon)/\pi\epsilon,\lambda_-,\lambda_+)U$, 
with $U$ and $\lambda_{\pm}$ given by Eqs.~\ref{App1-Eq:U},\ref{App1-Eq:lambda} in Appendix \ref{App1}.
The currents in terminal 3 (4) can be then written as
\begin{equation}
\langle I_{3(4),N}\rangle_{\delta}=\frac{e^2V}{h}\frac{1}{2}
\left(1\pm\frac{\lambda_+^Nu_+-\lambda_-^Nu_-}{u_+-u_-}\right),
\end{equation}
with $u_{\pm}$ given in Eq.~\ref{EqApp-upm} in Appendix \ref{App1}. 

\begin{figure}[t]
\begin{center}
Ê\includegraphics[width=8cm]{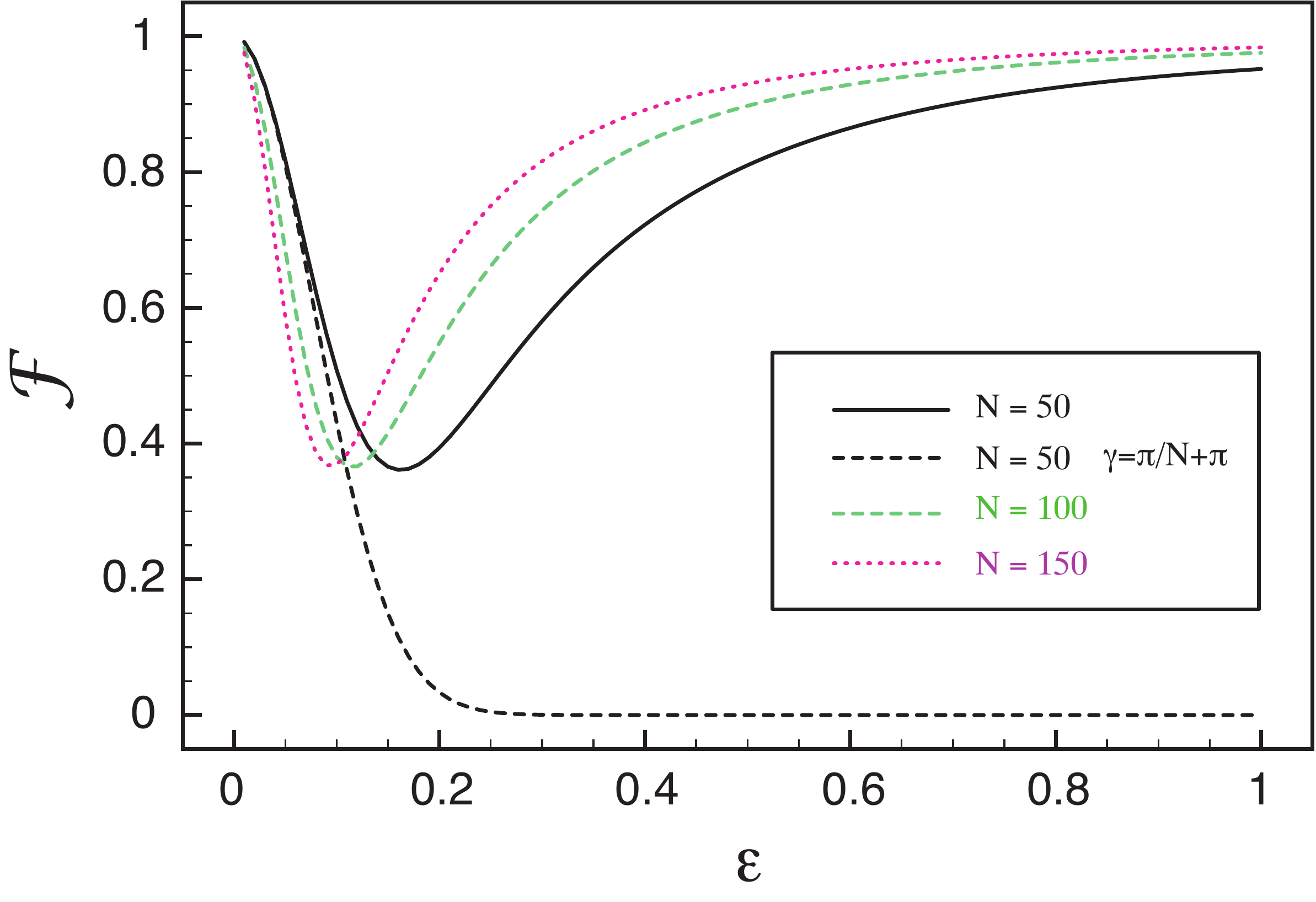}
ÊÊÊ\caption{Fraction of coherent signal ${\cal F}$ of Eq.~(\ref{Eq:F})
ÊÊÊversus the strength 
ÊÊÊ		$\epsilon$ of the fluctuating field. Choosing the degree of 
		admixture of the BSs to be $\gamma=\pi/N$, with most of the 
		electron amplitude injected in the coherent $i$ channel, the 
		outgoing signal initially partially dephases for small $\epsilon$, 				
		reaches a minimum, and then recovers its coherence as $\epsilon$ 			
		approaches one (IFM regime established). On the contrary, injecting most of the electron 				
		amplitude in the channel affected by random phase shift by 
		setting $\gamma=\pi/N+\pi$, the coherence of the outgoing signal is 			
		totally lost (no IFM regime). Ê\label{Fig3}}
\end{center}
\end{figure}

Figure~\ref{Fig2} (left panel) shows the current in terminal 3 Êversus the phase shift $\phi$ for the case of no dephasing ($\epsilon=0$).
We can see that for large $N$, $\langle I_{3,N}\rangle_{\delta}$ is approximately $e^2V/h$ for almost all values of $\phi$, and that only at 
$\phi=0$ it drops very rapidly to zero. For such value the outgoing currents are indeed
$\langle I_{3,N}\rangle_{\delta}=0$ and $\langle I_{4,N}\rangle_{\delta}=e^2V/h$ independently from $N$
(increasing $N$ further shrinks the dip at Ê$\phi=0$). 
This corresponds to having a very narrow resonance at the working point $\phi=0$ where
interference gives rise to a gradual transfer of the electron wave-function to the $o$ channel and all the 
current emerges from contact 4. Such a resonance is very sensitive to small deviations of the phase $\phi$ from the working 
point $\phi=0$ and imply a large variation of the current response.

In the Êcase of Êstrong dephasing ($\epsilon=1$) the current is instead given by Ê
\begin{eqnarray}
\langle I_{3(4),N}\rangle_{\delta}= \frac{e^2V}{h}\frac{1}{2}\left(
1\pm\cos^N (\gamma)
\right)\;.
\label{Eq:I34dephPhi=0}
\end{eqnarray}
If the asymmetry of the BSs is properly tuned at the value $\gamma=\pi/N$, the output currents are
$\langle I_{3(4),N}\rangle_{\delta}=\frac{e^2V}{h}\frac{1}{2}\left(
1\pm\cos^N\left(\frac{\pi}{N}\right)\right)$,
so that, in the limit of large $N$, one finds that $\langle I_{3,N} \rangle_\delta=e^2V/h$ and $\langle I_{4,N}\rangle_{\delta}=0$. The behavior of the current in contact 3 versus the dephasing strength $\epsilon$ is shown in Fig.~\ref{Fig2}, right panel. 
It is evident Êthat the presence of a strong dephasing source changes the interference response, 
so that for $N\gg1$ all electrons exit the device from terminal 3, whereas in the 
coherent case they would exit from terminal 4. Thus, in this respect the system
behaves like a ``which-path" electronic interferometer~\cite{ALE}. 
Interestingly we note that ÊEq.~(\ref{Eq:I34dephPhi=0}) predicts that, for even $N$, Ê
the same behavior can Êbe observed also in Êthe highly asymmetric case when the electronic amplitude is diverted to the noisy channel $o$, i.e. 
$\gamma = \pi/N+ \pi$. In the next section we shall see however that, differently to the case $\gamma = \pi/N$, Êthis last regime does not correspond to a true IFM effect, since the coherence of the transmitted signals is totally washed out.

\subsection{Coherence of the outgoing signal}

A key feature of the IFM detection of noise is that coherence 
of the output be preserved and this can open the way to novel applications in 
quantum-coherent electronics. Depending on whether the electron is mostly injected into the 
secure $i$ channel by setting $\gamma=\pi/N$ or into the $o$ channel exposed to dephasing, 
by setting $\gamma=\pi/N+\pi$, the coherence of the outgoing signal can be asymptotically preserved or totally lost.

An effective way to quantify the coherence of the outgoing signal can be obtained by defining the fraction of coherent signal as 
\begin{equation}\label{Eq:F}
{\cal F}\equiv|\langle t \rangle_{\delta}|^2+ 
|\langle r\rangle_{\delta}|^2 ,
\end{equation}
where we have set $t=[S_N]_{11}$ and $r=[S_N]_{21}$ so that
$\langle t \rangle_{\delta}$ ($\langle r \rangle_{\delta}$)
is the averaged transmission amplitude to contact 3 (4). ${\cal F}$ takes
values between 0 (complete loss of coherence) and 1 (coherence
fully preserved, since in this case $|[S_N]_{11}|^2+|[S_N]_{21}|^2=1$).
The two quantities $\langle t\rangle_{\delta}$ and $\langle r\rangle_{\delta}$
measure the coherence of the transmitted electrons into contacts 3 and 4, respectively, since they are proportional to the
interference terms of such electrons with a reference, coherent,
signal (a thorough discussion is given in Sec.~\ref{Sec3}).
In Fig.~\ref{Fig3} we plot ${\cal F}$ for different choices of $N$ and $\gamma$. 
For $\gamma=\pi/N$, the fraction of coherent signal initially decreases as a result of the disturbance induced by the fluctuating field (degradation of coherence).
For large values of $\epsilon$, however, the dephasing of the tiny portion of the wavefunction pertinent to the $o$ channel prevents the occurrence of destructive interference.
As a result Êfull, coherent transmission through the lower arm of the setup is established, yielding
${\cal F}\simeq 1$ and thus indicating that an IFM is taking place in the setup.
This can be understood as due to the quantum Zeno effect~\cite{ZENO} associated with repetitive measurements that try to determine whether or not Êthe electron is ``passing'' through the upper arm of the interferometer~\cite{NOI09}.
For $\gamma=\pi/N$, the outcome of such a measurement will be negative with a very high probability (i.~e.~the electron is
found in the lower arm) preserving coherence. An interplay between these two regimes occurs for intermediate
values of $\epsilon$ giving rise to a minimum in ${\cal F}$ which sharpens for higher $N$ (see Fig.~\ref{Fig3}).
This scenario changes completely for $\gamma=\pi/50+\pi$. Here electrons are mostly injected into the $o$ channel. For small values of $\epsilon$ the situation is analogous to the case $\gamma=\pi/N$, the behavior of ${\cal F}$ being actually the same: the noise source induces a partial suppression of the destructive interference yielding a consequent degradation of coherence.
As evident from Fig.~\ref{Fig3} however, in this case large values of $\epsilon$ yields a drop of ${\cal F}$ to zero indicating that 
no IFM is taking place here.
This originates from Êthe fact that Êthe complete suppression of the destructive interference is accompanied by a likewise complete loss of coherence due to the strong dephasing experienced by the electron.
Ê

\begin{figure}[t]
\begin{center}
Ê\includegraphics[width=8cm]{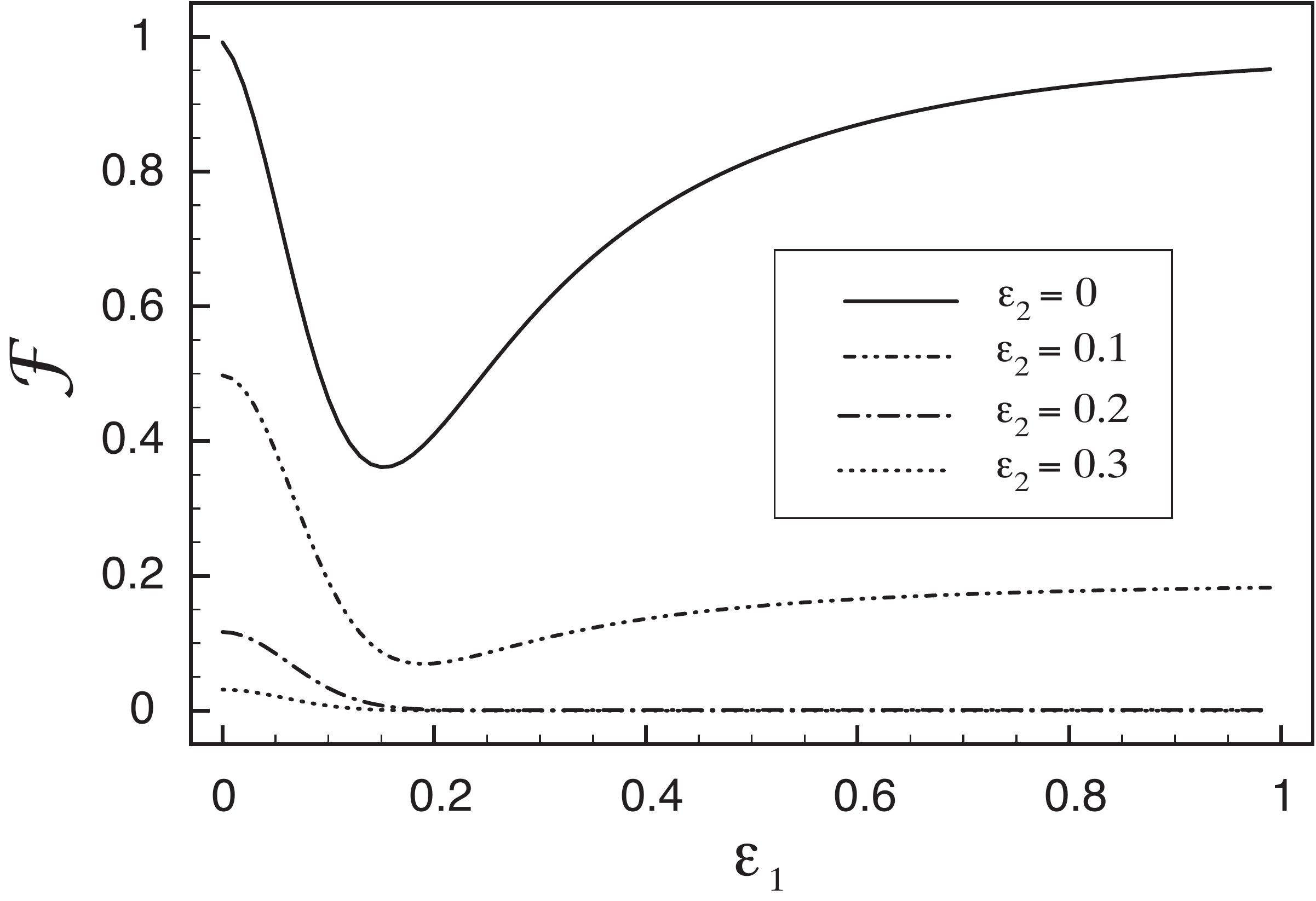}
ÊÊÊ\caption{Fraction of coherent signal ${\cal F}$ when the dephasing field affects both Ê		
ÊÊÊ		channels, respectively $o$ with strength $\epsilon_1$ and $i$ with 			
		strength $\epsilon_2$. The degree of admixture is set to $\gamma=			
		\pi/N$, with $N=50$, and most of the electron amplitude is injected in 		
		channel $i$. By increasing $\epsilon_2$ the coherence is rapidly 			
		lost. \label{Fig4}}
\end{center}
\end{figure}

So far we have considered an ideal situation in which dephasing takes place only in the $o$ 
channel. Figure~\ref{Fig4} shows the behavior of ${\cal F}$ versus the strength $\epsilon_1$ 
of the dephasing field acting on channel $o$, when a fluctuating field of strength $\epsilon_2$ 
affects propagation in the $i$ channel. We see that a strong response corresponds to a slight 
increase of $\epsilon_2$, with the coherence of the outgoing signal being significantly degraded. Ê

\subsection{Detection of the coherent signal}
\label{Sec3}

\begin{figure}[t]
\begin{center}
Ê\includegraphics[width=8cm]{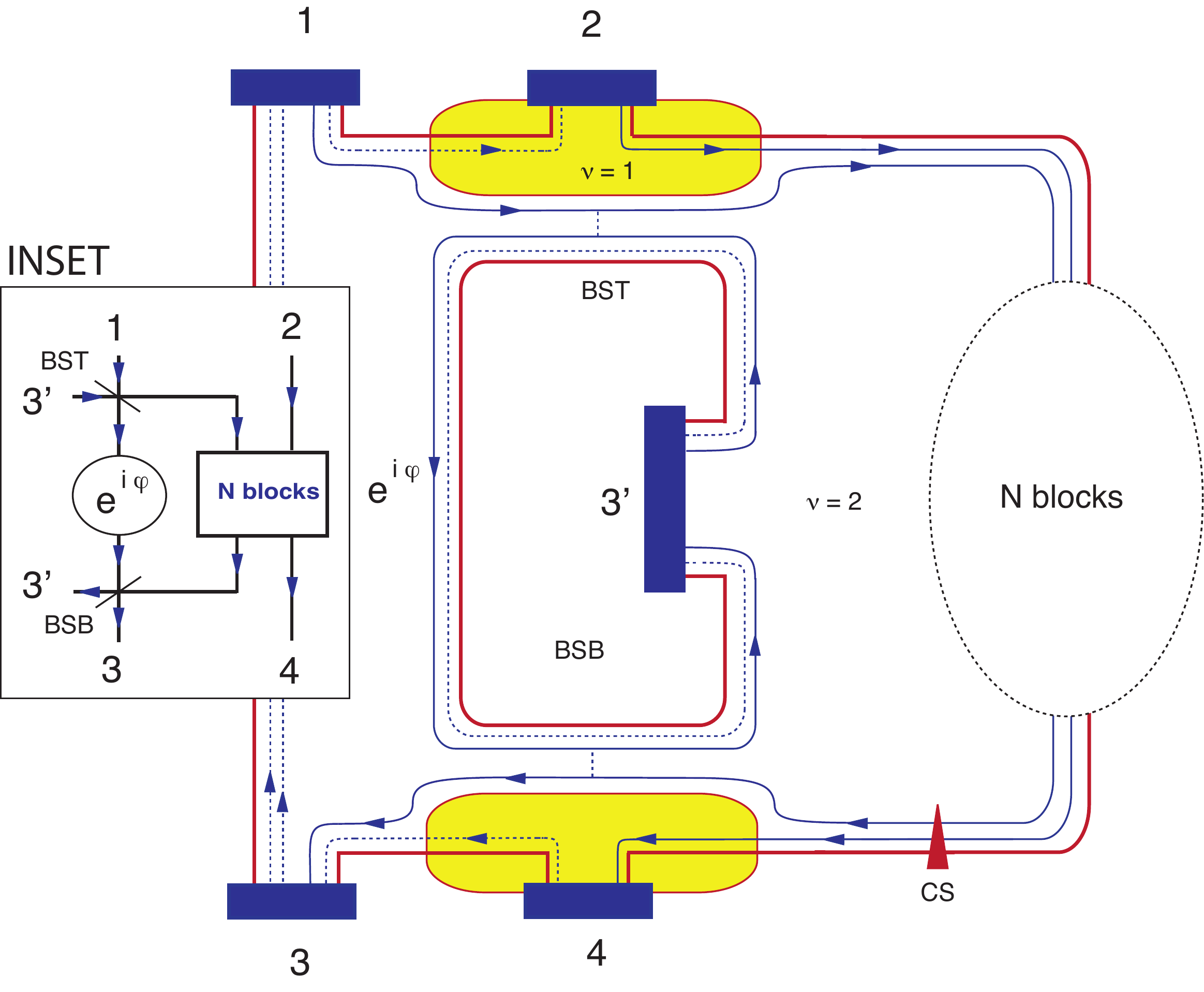}
ÊÊÊ\caption{Schematic representation of the proposal for an experimental 				
ÊÊÊrealization of an $N$-block noise-sensitive electron 						
ÊÊÊchannel embedded in a Mach-Zehnder interferometer. Electrons 			
ÊÊÊentering the a Hall bar from contact 1 split at the beam splitter BST. Ê			
ÊÊÊThe electrons transmitted will traverse the $N$-block system and 			
ÊÊÊeventually go out from contact 4 or impinge onto BSB. The latter mix 			
ÊÊÊwith those initially reflected at BST and interfere. The result of the 			
ÊÊÊinterference can be collected in contact 3 or 3'. ÊIn the yellow areas 			
ÊÊÊthe filling factor is $\nu=1$ and in the rest of the Hall bar the filling 			
ÊÊÊfactor is $\nu=2$. The coherence of the outgoing signal can be 				
ÊÊÊdirectly addressed by measurement of the visibility of 					
ÊÊÊcurrent in contact 3 versus the tunable phase $\varphi$ 					
ÊÊÊacquired during the propagation by the electron reflected at BST. 			
ÊÊÊInset: Schematics of the main picture. \label{Fig5}}
\end{center}
\end{figure}

In this section we show that the fraction of coherent signal ${\cal F}$ defined in Eq.~\ref{Eq:F} can actually be measured by
embedding the $N$ concatenated blocks in a Mach-Zehnder interferometer, as schematically illustrated in Fig.~\ref{Fig5}.
A voltage $V$ is applied to contact 1, while all other contacts are at reference potential.
A beam splitter (BST in Fig.~\ref{Fig5}) splits the current injected by contact 1 so that the transmitted portion
enters the $N$-block system from channel $i$, while the reflected one follows a path whose length (and phase $\varphi$) 
can be arbitrarily adjusted. 
The current exiting the $N$-block system via channel $i$ is then mixed with the signal of known phase at beam splitter BSB. The two outgoing currents are collected by contacts 3 and 3'. Electrons exiting the $N$-block system from channel $o$ are drained separately by contact 4.

\begin{figure}[t]
\begin{center}
Ê\includegraphics[width=8cm]{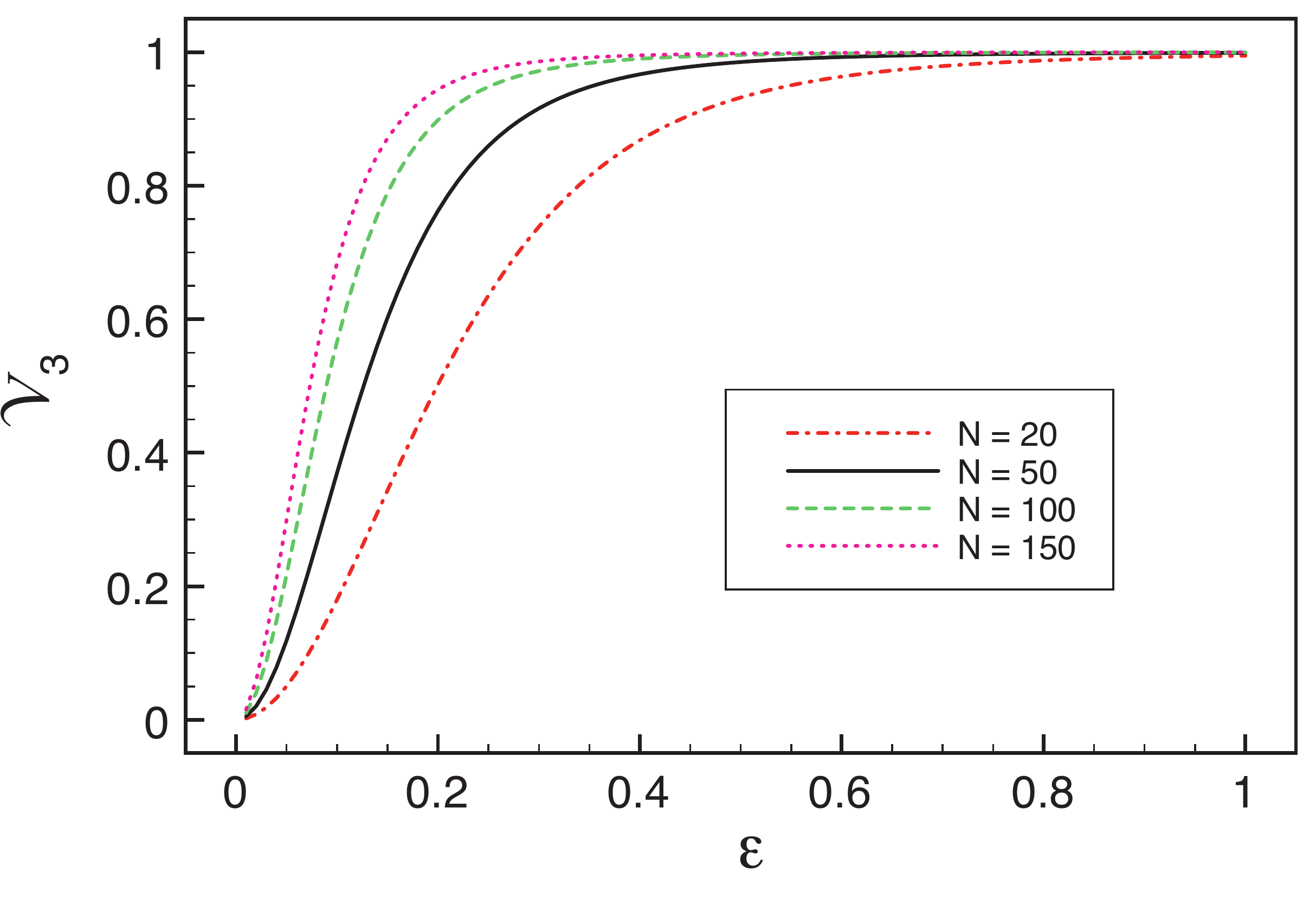}
ÊÊÊ\caption{Visibility~(\ref{viz}) of the current in contact 3 versus the strength of the 
ÊÊÊdephasing field, for several numbers of blocks $N$. In the coherent case $\epsilon=0$ 
ÊÊÊthe current in contact 3 is zero and so is ${\cal V}_3$. Increasing $\epsilon$ the 
ÊÊÊvisibility approaches one. We set $\gamma=\pi/N$. \label{Fig6}}
\end{center}
\end{figure}

Assuming that both BST and BSB are 50/50 beam splitters, the transmission probability for electrons to exit via contact 3 is given by
\begin{eqnarray}
{\cal T}_3(\varphi)&=&\frac{1}{4}\left\langle\left|t+e^{i\varphi}\right|^2\right\rangle_{\delta} \nonumber \\
&=&\frac{1}{4}(\left\langle T\right\rangle_{\delta}+1)+\frac{1}{2}
|\left\langle t\right\rangle_{\delta}|\cos\left({\rm arg}
(\left\langle t\right\rangle_{\delta})-\varphi\right)\;, \nonumber 
\end{eqnarray}
where we recall that $t$ is the amplitude for electrons to exit from the $N$ concatenated interferometers Ê
in channel $i$, and $T=|t|^2$. 
The visibility of ${\cal T}_3(\varphi)$ is defined as the maximal normalized amplitude of the $\varphi$-oscillation, namely
\begin{equation}
{\cal V}_3=\frac{2|\left\langle t\right\rangle_{\delta}|}
{\left\langle T\right\rangle_{\delta}+1}.
\label{viz}
\end{equation}
Figure~\ref{Fig6} shows function ${\cal V}_3$ versus $\epsilon$ with $\gamma=\pi/N$, for different numbers of interferometers ($N$). At $\epsilon=0$ the destructive interference for $\phi=\pi$ produces a zero amplitude 
signal $t$, leading to zero visibility. In the presence of the dephasing field the visibility rapidly increases and saturates to one, thereby revealing the coherence of the amplitude $t$ with respect to the phase $\varphi$. 

Analogously, transmission probability ${\cal T}_4$ is related to the amplitude $r$ of electrons exiting from the $N$ concatenated interferometers from channel $o$. ${\cal T}_4$ can be measured by tuning the beam splitter CS in Fig.~\ref{Fig5} in order to swap inner and outer channels.
One finds that Ê
\begin{equation}
{\cal T}_4(\varphi)=\frac{1}{2}(\left\langle R\right\rangle_{\delta}+1)+
|\left\langle r\right\rangle_{\delta}|\cos\left({\rm arg}
(\left\langle r\right\rangle_{\delta})-\varphi\right),
\end{equation}
with $R=|r|^2$, and visibility ${\cal V}_4$ is defined analogously to (\ref{viz}).
If we label $\bar{\cal T}_3$ ($\bar{\cal T}_4$) the mean value with respect to the phase of the transmission probability in 3 (4), we can write
\begin{equation}\label{Eq:FvsV}
{\cal F}={\cal V}_3^2\bar{\cal T}_3^2+{\cal V}_4^2\bar{\cal T}_4^2.
\end{equation}

In order to allow only a small fraction of the electron wavefunction to propagate in the dephasing $o$ channelÊand realize the conditions that allow IFMs, it is necessary to set the degree of admixture in the BS to the precise value 
$\gamma=\pi/N$. This may represent a technical obstacle to an experimental realization, since BSs are 
difficult to be tuned all to the same precise degree of admixture, and a high efficiency IFM is obtained in the limit of large $N$.
In the following we shall present a more robust architecture that allows one to overcome this difficulty by translating the spatial concatenation to the time-domain regime.

\section{Multiple interference in the time domain}
\label{Sec4}

In this section we show that it is possible to implement an IFM scheme based on the integer-quantum Hall Mach-Zehnder (MZ) interferometer of the type experimentally realized in Refs.~[\onlinecite{JiNature03,NederPRL06,LitvinPRB07,RoulleauPRB07,NederNatPh07,SamuelssonPRL04,NederNatur07}] by exploiting a quantizing electron 
emitter~\cite{Gabelli06, Feve07,Moskalets08, Splettstoesser08}. 

\begin{figure}[t]
\begin{center}
Ê\includegraphics[width=8cm]{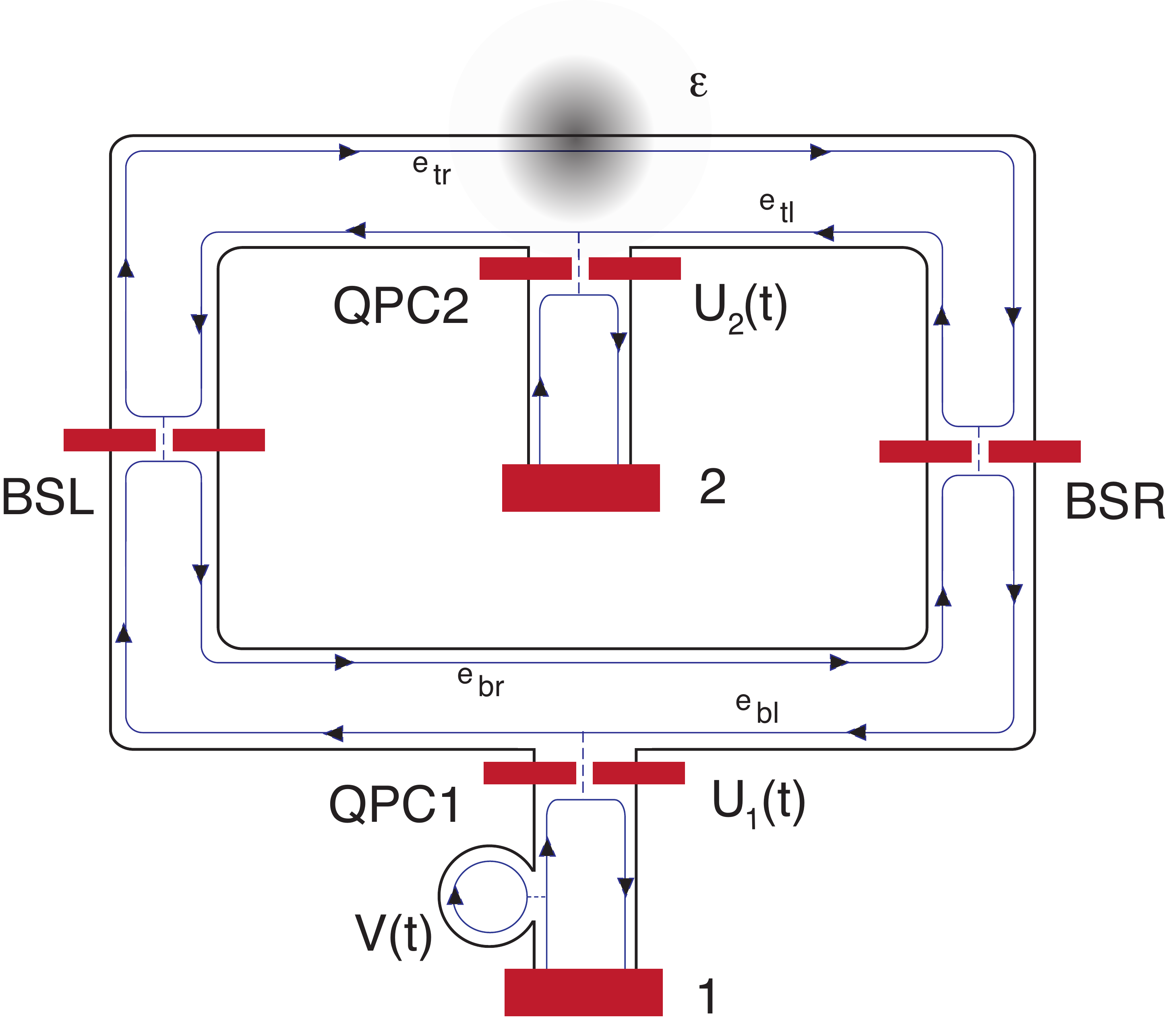}
ÊÊÊ\caption{Mapping of concatenation in space to the time domain in a Mach-			
ÊÊÊZehnder interferometer. A time-dependent voltage generates a 				
ÊÊÊcurrent of well separated electron and holes. The QPC1 let the 				
ÊÊÊelectrons enter the Mach-Zehnder, perform $N$ rounds in the 				
ÊÊÊinterferometer and then collect them back into contact 1 in the case a 		
ÊÊÊdephasing field of strength $\epsilon$ affects the dynamics of the 			
ÊÊÊchannel $e_{tr}$. In the coherent case $\epsilon=0$ Êthe electrons are collected in 			
ÊÊÊcontact 2. Ê\label{Fig7}}
\end{center}
\end{figure}

Figure~\ref{Fig7} shows a schematic view of the MZ interferometer, which comprises two beam splitters, two electrodes coupled through quantum point contacts (QPC1 and QPC2), and a dephasing source affecting the propagation of electrons in the edge channel $e_{tr}$.
A small weakly-coupled circular cavity is placed between contact 1 and QPC1. This produces a train of time-resolved electron and hole wave packets (details of such single electron source can be found in Appendix~\ref{ehswitch}).
Every period comprises a pair of electron and hole pulses, as shown in Fig.~\ref{Fig10}.
QPC1 and QPC2 are controlled by the time-dependent external potentials $U_1(t)$ and $U_2(t)$.

The system is operated as follows.
In the first period, QPC1 is opened during the first half cycle letting the electron pulse to be injected into the MZ. It is closed during the second half so that holes will be reflected back into lead 1.
The injected electron propagates Êwith velocity $v_F$ along the edge $e_{bl}$ of the MZ until it meets the first beam splitter BSL where it is split into two packets that follow two different edge channels ($e_{tr}$ and $e_{br}$) of equal length $L$ and finally reach the second beam splitter BSR after a time $L/v_F$.
Here the two packets interfere and then propagate along edges $e_{bl}$ and $e_{tl}$ of length $L$.
Keeping QPC1 and QPC2 closed, the sequence repeats itself with the electronic wavepacket being split and reunited many times at beam splitters BSL and BSR. This propagation is fully equivalent to a spatial concatenation of distinct MZ interferometers.
At a chosen time, the electron pulse can be collected from leads 1 and 2 by opening QPC1 and QPC2, respectively.

Let us assume that an electron at time $t_+$ and a hole at time $t_-$ arrive at QPC1, with $0\leq t_+\leq{\cal T}/2$ and ${\cal T}/2\leq t_-\leq{\cal T}$, ${\cal T}$ being the period of the cycle.
The electron injected through QPC1 at time $t_+$ will appear at one of the two QPCs after 
a time $t_++N\Delta t$, with $\Delta t\equiv2L/v_F$, after performing $N$ rounds.
The two QPCs are then opened simultaneously. 
In the case where no dephasing field is present, $\epsilon=0$, it is possible to tune the MZ such that after $N$ rounds 
the electron pulse is at QPC2 and can be collected in contact 2.
In the case of maximal dephasing, $\epsilon=1$, the electron pulse is at QPC1.

Energy-level spacing inside the MZ can be estimated as $\Delta E\sim h/\Delta t$.
$L$ can be chosen to be large enough for a continuum approximation of the level spacing to be valid.
This picture allows us to describe the physics in the Landauer-B\"uttiker formulation, with no needs of the Floquet treatment of this time-dependent problem. We introduce the electron annihilation operators Ê$\{\hat{e}_{tr}, \hat{e}_{br}, \hat{e}_{bl}, \hat{e}_{tl}\}$ that annihilate an electron on the edge states $\{e_{tr}, e_{br}, e_{bl}, e_{tl}\}$. 
In order to obtain the transport regime described in the previous section we must tune beam splitters BSL and BSR so that 
\begin{equation}
S_{BSL}=S_{BSR}=\left(\begin{array}{cc}
\cos(\gamma/2) & i\sin(\gamma/2)\\
i\sin(\gamma/2) & \cos(\gamma/2)
\end{array}\right), 
\end{equation}
with $(\hat{e}_{tr}, \hat{e}_{br})^T=S_{BSL}(\hat{e}_{bl}, \hat{e}_{tl})^T$ and $(\hat{e}_{bl}, \hat{e}_{tl})^T=S_{BSR}(\hat{e}_{tr}, \hat{e}_{br})^T$, 
with the particular choice $\gamma=\pi/N$.
Concerning the dynamical phase acquired by propagating along the edge channels, 
arms of equal length $L$ do not give rise to a relative phase shift, and the condition for the working point $\phi=0$ depends only 
on the applied magnetic-field intensity. 

\section{IFM with an Aharonov-Bohm ring}
\label{Sec5}

\begin{figure}[t]
\begin{center}
Ê\includegraphics[width=8cm]{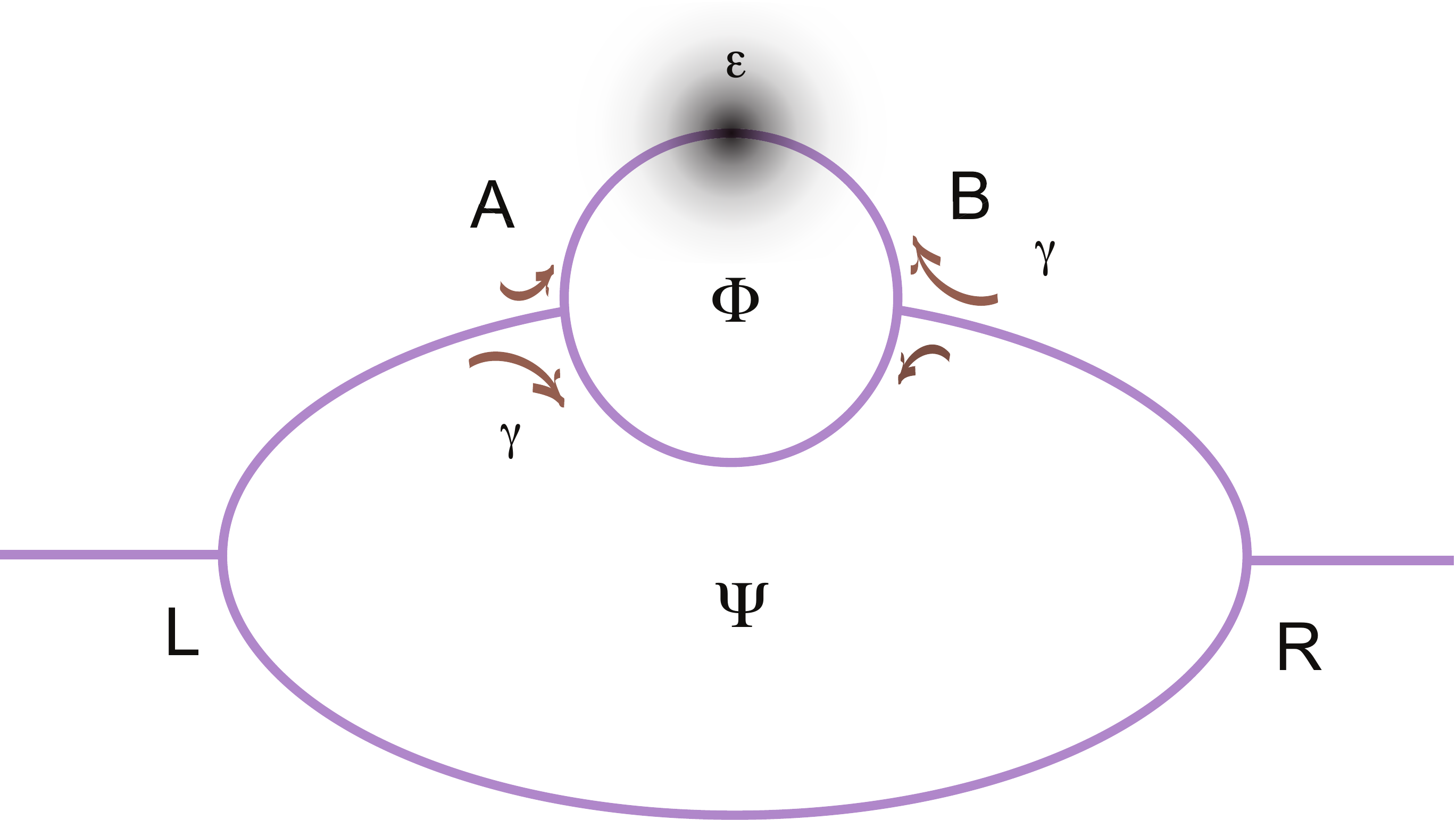}
ÊÊÊ\caption{Schematic representation of a double ring setup that allow to quantify 		
ÊÊÊvia a current measurement the degree of coherence of the signal 			
ÊÊÊgoing out from the small ring. The larger ring is pierced by a magnetic 		
ÊÊÊflux $\Psi$ and the small ring by a flux $\Phi$. The nodes $L$ and 			
ÊÊÊ$R$ of large ring split the electron amplitude impinging on them in 			
ÊÊÊthe two arms of the large ring in a symmetric way, whereas The 				
ÊÊÊnodes $A$ and 	$B$ of small ring split the electron amplitude 				
ÊÊÊimpinging on them in the two arms of the small ring in a non 				
ÊÊÊsymmetric way according to the parameter $\gamma$. ÊA dephasing 			
ÊÊÊfield of strength $\epsilon$ affects the dynamics of electrons traveling 		
ÊÊÊin the upper arm of the small ring by randomly shifting their phase.\label{Fig8}}
\end{center}
\end{figure}

In this section we review the implementation of the IFM scheme using an asymmetric Aharonov-Bohm (AB) ring proposed in Ref.~[\onlinecite{NOI09}] 
and discuss a scheme allowing the direct test of output-signal coherence. 
This latter task can be performed by embedding the asymmetric AB ring in a larger, symmetric AB ring. We shall examine the case in which the smaller ring is placed in the upper arm of the larger one, as shown in Fig.~\ref{Fig8}.

We shall use again the Landauer-B\"uttiker formalism of quantum transport and assume that the small asymmetric AB ring supports a single channel.
Following Ref.~[\onlinecite{NOI09}], we parametrize the scattering matrix connecting the incoming to the outgoing modes in node $A$ as 
\begin{equation}
S_A = \left(\begin{array}{cc}
r_A & \bar{\bf t}_A\\
{\bf t}_A & \bar{\bf r}_A \end{array}\right)
=\left(\begin{array}{ccc}
a & b \cos(\frac{\pi}{2}\gamma) & b \sin(\frac{\pi}{2}\gamma) \\
b \sin(\frac{\pi}{2}\gamma) & a & b \cos(\frac{\pi}{2}\gamma) \\
b \cos(\frac{\pi}{2}\gamma) & b \sin(\frac{\pi}{2}\gamma) & a\\
\end{array}\right) \nonumber 
\end{equation}
with $r_A=a$, ${\bf t}_A$ the $2\times 1$ bottom left block, Ê$\bar{\bf t}_A$ the 
$1\times 2$ top right block and Ê$\bar{\bf r}_A$ the remaining $2\times 2$ bottom 
right block, with $a=-\sin(\pi\gamma)/(2+\sin(\pi\gamma))$ and $b=\sqrt{1-a^2}$. Similarly, for node $B$ 
\begin{equation}
S_B=\left(\begin{array}{cc}
\bar{r}_B & {\bf t}_B\\
\bar{\bf t}_B & {\bf r}_B \end{array}\right).
\end{equation}
We further assume injection invariance under node exchange. This configuration was theoretically 
studied and experimentally realized at low magnetic fields\cite{SzafranPRB05,SzafranEu05,Strambini09} and can be 
understood as the result of Lorentz force. We label annihilation operators for incoming $(L)$ and 
outgoing $(u,d)$ modes in node $A$ as ${\bf a}_L\equiv(a_L,a_u,a_d)^T$ and ${\bf b}_L\equiv(b_L,b_u,b_d)^T$ 
respectively, so that ${\bf b}_L=S_A{\bf a}_L$. Analogously we label incoming and outgoing 
modes in node $B$ as ${\bf a}_R\equiv(a_R,a'_u,a'_d)^T$ and ${\bf b}_R\equiv(b_R,b'_u,b'_d)^T$, respectively, 
with ${\bf b}_R=S_B{\bf a}_R$. Symmetry under cyclic exchange of nodes $A$ and $B$ implies that
\begin{equation}
\left(\begin{array}{c}
b_R \\ b'_d \\ b'_u
\end{array}\right)=S_A
\left(\begin{array}{c}
a_R \\ a'_d \\ a'_u
\end{array}\right).
\end{equation}
By rearranging the order of the vector components we obtain $S_B=S_A^T$. 
$\gamma$ controls the asymmetry of nodes $A$ and $B$, so that for $\gamma=0$ ($\gamma=1$) complete 
asymmetry is achieved, with the electron entering from the left lead being injected totally in the lower (upper) arm, whereas for $\gamma=1/2$ the injection is symmetric. 
An external magnetic field is applied perpendicularly to the plane and is 
responsible for the magnetic Aharonov-Bohm phase acquired in the ring. 
At the same time it yields the Lorenz force which leads to the ring asymmetry.
Electron propagation in the two arms is described by matrices $S_p(\delta) = e^{i k_F \ell} \;\mbox{diag}(e^{i\phi/2+i\delta}, e^{-i \phi /2})$, for transmission from left to right, and $\bar{S}_p(\delta)= e^{i k_F \ell} \;\mbox{diag}(e^{-i\phi/2+i\delta}, e^{i\phi/2})$, for transmission from right to left.
Here $\phi$ is the ratio of the magnetic-field flux through the asymmetric ring to the flux quantum, $k_F$ is the Fermi wavenumber, $\ell$ is 
the length of the arms and $\delta$ is an additional random phase. In the following we shall set $k_F \ell=\pi/2$ and anticipate that a different choice does not change qualitatively our findings.

As mentioned earlier, the asymmetric AB ring is embedded in a larger symmetric AB ring so that the phase that an electron accumulates while traveling in the lower arm of the large ring represents a reference for the electron that traverses the asymmetric ring.
By tuning the magnetic field that pierces the larger ring, we can determine the visibility of the current which reflects the loss of coherence occurring in the small asymmetric ring.

We describe scattering at nodes $L$ and $R$ of the large ring by a scattering matrix \cite{Buttiker84}
\begin{equation}
S_L=
\left(\begin{array}{cc}
r_L & \bar{\bf t}_L\\
{\bf t}_L & \bar{\bf r}_L \end{array}\right)=
\left(\begin{array}{ccc}
c & \sqrt{g} & \sqrt{g}\\
\sqrt{g} & d & e\\
\sqrt{g} & e & d
\end{array}\right),
\end{equation}
with $r_L=c$, ${\bf t}_L$ the $2\times 1$ bottom left block, Ê$\bar{\bf t}_L$ the 
$1\times 2$ top right block and Ê$\bar{\bf r}_L$ the remaining $2\times 2$ bottom 
right block. The scattering matrix depends only on parameter $g$, which controls the lead-to-ring coupling strength via $c=\sqrt{1-2g}$, $d=-(1+c)/2$, $e=(1-c)/2$, with $0<g<1/2$.
On the right node we have $S_R=S_L^{\dag}$.
Free propagation along the large-ring arms (assumed to be of equal length $L$) is accounted for by splitting the ring into two halves, each of which is described by the $2 \times 2$ diagonal matrix $P=e^{ik_FL/2}{\rm diag}(e^{i\varphi/4},e^{-i\varphi/4})$, for propagation from left to right, and $\bar{P}=e^{ik_FL/2}{\rm diag}(e^{-i\varphi/4},e^{i\varphi/4})$, for propagation from right to left. Here $\varphi$ Êis the ratio of the magnetic-field flux through the larger symmetric ring ($\Psi$) to the flux quantum.
The overall amplitude for transmission $\tau$ from the left to the right lead is calculated through a multiple scattering formula which takes into account all interference processes between possible paths that electrons can take to go from the left to the right.
In the absence of decoherence one finds (see Appendix~\ref{App:TrRef}):
\begin{equation}
\tau=\boldsymbol{\tau}_B(\openone-\Gamma)^{-1}S'_p\boldsymbol{\tau}_A\;.
\end{equation}

\subsection{Transmission in the presence of a dephasing field}

We now assumeÊa fluctuating external field (dephasing source) is placed in the upper arm of the small asymmetric ring.
This can be described by defining the partial transmission amplitude of order $N$ with
\begin{equation}
t_N=\boldsymbol{\tau}_B\sum_{n=0}^N\prod_{j=0}^n\Gamma_{(n-j)}S'_{p,0}
\boldsymbol{\tau}_A ,
\end{equation}
where $\Gamma_{(j)}\equiv\Gamma(\delta_j,\delta_j') = 
S'_p(\delta_j)\bar{\boldsymbol{\rho}}_A\bar{S}'_p(\delta_j')\boldsymbol{\rho}_B$ depends on two random phases $\delta_j$ and $\delta'_j$, and $S'_{p,0}\equiv S'_p(\delta_0)$. 
As in Sec.~\ref{Sec2}
we then choose the random phases from a uniform distribution 
of zero mean and width $2\pi\epsilon$ and Êcompute Êthe averaged 
partial transmission probability as $\langle t_N^* t_N \rangle_{\delta}$.
It can be shown that the following recursive relation holds:
\begin{equation}
\langle t_N^*t_N\rangle_{\delta}=\langle t_{N-1}^*t_{N-1}
\rangle_{\delta}+\Xi_N.
\end{equation} 
By iterating the procedure, the averaged transmission probability $\langle T\rangle_{\delta} = \lim_{N\rightarrow\infty} \langle t_N^* t_N\rangle_{\delta}$
can be written as 
$\langle T\rangle_{\delta}=\sum_{N=0}^{\infty}\Xi_N$.
To compute such limit we introduce Êthe Gell-Mann matrix vector $\boldsymbol{\Sigma}=
(\Sigma_0,\Sigma_1,\ldots,\Sigma_8)^T$, with $\Sigma_0=\sqrt{2/3}\times\openone$, 
write $\boldsymbol{\tau}^{\dag}_B\boldsymbol{\tau}_B={\bf p}_B\cdot\boldsymbol{\Sigma}$, with $({\bf p}_B)_i=\frac{1}{2}{\rm Tr}(\boldsymbol{\tau}^{\dag}_B\boldsymbol{\tau}_B\Sigma_i)$, and define the following decoherence matrix 
\begin{equation}\label{Eq:DecohMap2}
{\cal Q}_{ij}=\frac{1}{2}
\langle {\rm Tr}\left[
\Gamma^{\dag}({\delta})\Sigma_i\Gamma({\delta})
\Sigma_j\right]\rangle_\delta ,
\end{equation}
which allows us to perform the average over the random phase as a matrix product. 
Similarly we define $\Gamma_{\rm av}=
\langle \Gamma(\boldsymbol{\delta})\rangle_\delta$,
and the decoherence map ${\cal P}$ with entries
\begin{equation}
{\cal P}_{ij}=\frac{1}{2}
\langle {\rm Tr}\left[
S_p^{\dag}(\delta)\Sigma_i S_p(\delta)\Sigma_j\right]
\rangle_\delta,
\end{equation}
that describes the average over the random phase in $S'_{p,0}$. Ê
$\Xi_N$ Êcan be concisely written as 
\begin{equation}
\Xi_N=\left({\bf p}_B\cdot {\cal Q}^N
+\sum_{k=1}^N{\bf p}_k\cdot{\cal Q}^{N-k}\right)
\cdot{\cal P}\cdot~\boldsymbol{\tau}_A^{\dag}~\boldsymbol{\Sigma}~
\boldsymbol{\tau}_A,
\end{equation}
with the vector $({\bf p}_k)_i=\frac{1}{2}\left[{\rm Tr}(\boldsymbol{\tau}^{\dag}_B
\boldsymbol{\tau}_B\Gamma_{\rm av}^k\Sigma_i)+c.c\right]$. 
By writing ${\bf p}_k={\rm Re}[\lambda_1^k\Lambda_1+
\lambda_2^k\Lambda_2+\lambda_3^k\Lambda_3]
\cdot{\bf p}_B$, with $\lambda_i$ the eigenvalues of $\Gamma_{\rm av}$, 
$U$ the matrix of the eigenvectors of $\Gamma_{\rm av}$, and 
$(\Lambda_i)_{jk}=(U\Sigma_j\Sigma_kU^{-1})_{ii}$,
that satisfy $(\Lambda_1+\Lambda_2+\Lambda_3)/2=\openone$, we can perform Ê
the sum on $N$ obtaining
\begin{eqnarray}\label{Eq:Dephased-T}
\langle T\rangle_{\delta}&=&{\bf p}_B\cdot
({\cal T}-\openone)\cdot(\openone-{\cal Q})^{-1}
\cdot{\cal P}\cdot~\boldsymbol{\tau}^{\dag}_A~\boldsymbol{\Sigma}~
\boldsymbol{\tau}_A, 
\end{eqnarray} 
with ${\cal T}$ being a $9\times 9$ matrix defined by
${\cal T}=\sum_{i=1}^3{\rm Re}[(1-\lambda_i)^{-1}\Lambda_i^T].$
The averaged transmission probability $\langle T\rangle_{\delta}$ is now function of the AB phase $\varphi$.

\subsection{Current as a measure of coherence}

The coherence of the signal transmitted through the small, asymmetric AB ring can be established by studying the
transport properties of the entire device. 

We focus on the case of strong coupling ($g\lesssim 1/2$) for which an electron approaching the large ring from node $L$ is mostly transmitted into the two arms of the large ring ($g=0.49$ in the following.)
For clarity, we also set the magnetic field and the arm length so that 
$\phi=\pi$, $\varphi=0$, $k_F\ell=\pi$, and $k_FL=\pi$. 
Actually, in a realistic experimental implementation it would be difficult to realize such conditions. We note however that the degree of coherence could be studied by changing one of the parameters of the large ring (e.~g.~$k_FL$) and measuring the visibility of the oscillations of the output signal.

\begin{figure}[t]
\begin{center}
Ê\includegraphics[width=8cm]{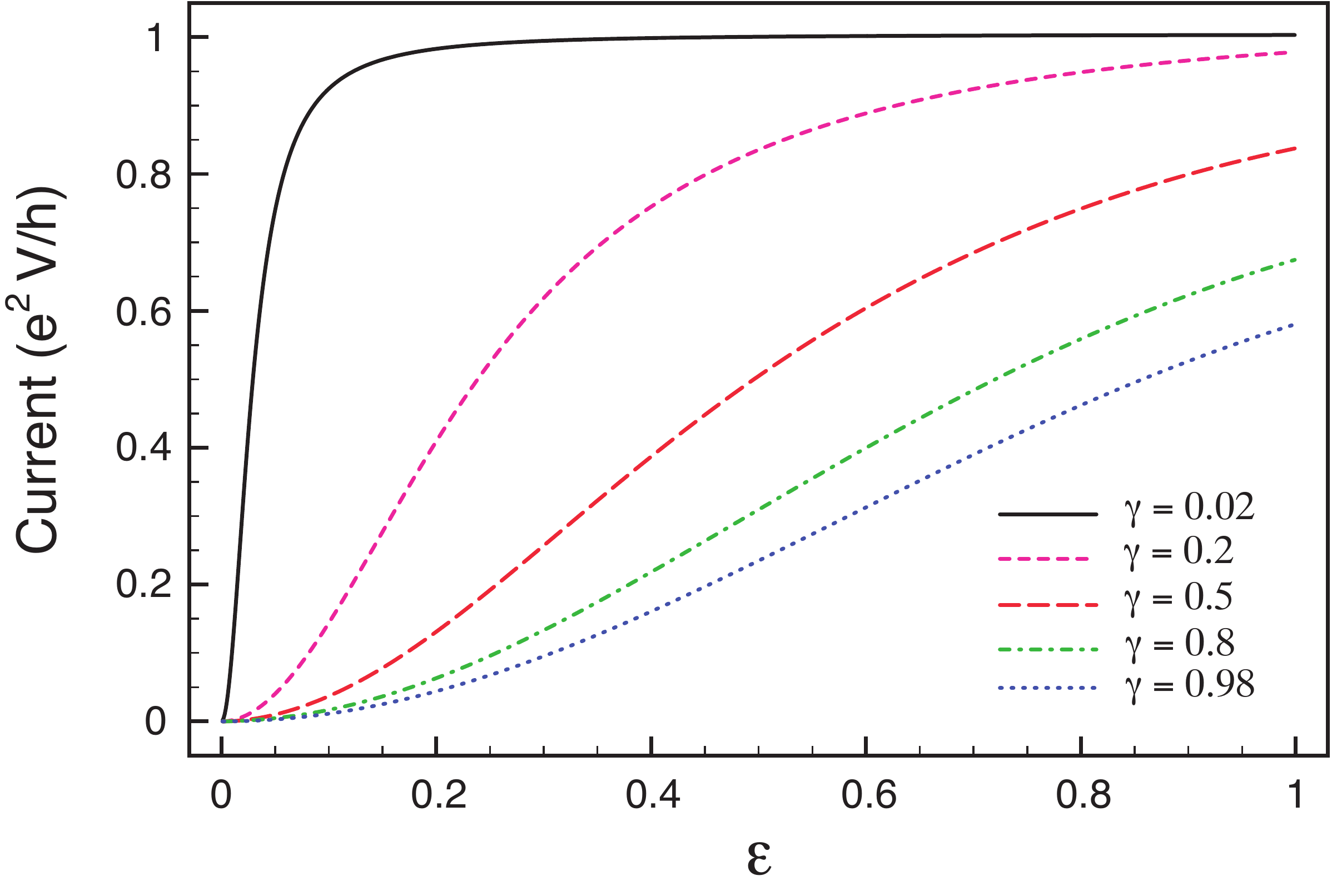}
ÊÊÊ\caption{Plot of the Êcurrent~(\ref{current1}) in units of $e^2V/h$, flowing from the left lead to the right lead of the double ring structure represented in Fig.~\ref{Fig8}, versus the strength $\epsilon$ of the dephasing field, at several degree of asymmetry $\gamma$. For $\gamma\rightarrow 1$ we divert the electrons mostly toward the dephasing source and consequently we have a reduction of the current flowing in the device. For $\gamma\rightarrow 0$ we divert the electron mostly toward the dephasing-free region Êand the coherent propagation gives rise to a maximal current flowing in the device. ÊPlot realized with $g=0.49$, $\phi=\pi$, $\varphi=0$, $k_F\ell=\pi$ and $k_FL=\pi$.
ÊÊÊ\label{Fig9}}
\end{center}
\end{figure}

For an applied bias voltage $V$, the zero-temperature current Êthrough the device of Fig.~\ref{Fig8} is given by
\begin{equation}
\label{current1}
I=\frac{e^2V}{h}\langle T\rangle_{\delta}\;,
\end{equation}

with $\langle T\rangle_{\delta}$ as in Eq.~\ref{Eq:Dephased-T}.
Figure~\ref{Fig9} shows $I$ Êas a function of the noise parameter $\epsilon$ for various values of the small-ring asymmetry parameter $\gamma$. 
As the dephasing strength $\epsilon$ is increased, however, $I$ increases with a behavior that strongly depends on the degree of asymmetry of the small ring. 
In the case of maximum decoherence ($\epsilon=1$) two different cases can be distinguished.
For $\gamma=0.02$ most of the electron amplitude that enters the small ring from the left will propagate into the lower arm of the small ring and coherently transmit into node R. There it interferes constructively with the reference path, saturating the current to the maximum $e^2V/h$. 
On the other hand for $\gamma=0.98$ most of the electron amplitude that enters from the left into the small ring will propagate into the upper arm of the small ring. There a dephasing field is present and the signal that propagates through the small ring will combine at node R with the reference path. The current exiting the device reaches a maximal value between zero and $e^2V/h$.
We interpret this behavior as an IFM of the dephasing field. The current exiting the device is proportional to the visibility of the output signal of the small asymmetric ring.

\section{Conclusion}

Based on the idea first suggested in Ref.~[\onlinecite{NOI09}] and directly inspired to the original proposal of A. Eliztur and L. Vaidman [\onlinecite{ElitzurVaidman}], 
in this article we focused on studying and detecting the presence of a classical external random fluctuating electric or magnetic field, which represent a common dephasing source in quantum devices. 
The noise source randomizes the phase of a propagating electron and plays the role of absorption in optical schemes while the loss of coherence of the outgoing electrons mimics photon absorption. The fraction of coherent output signal or alternatively the visibility of the outgoing signal Êrepresent the figures of merit that qualify an IFM. The study of this quantities allowed us to point out the difference between a ``which-path" detection and an IFM: the former allows only the detection of the presence of a dephasing source at the expense of the degradation of the visibility of the outgoing signal, whereas the latter allows a coherent detection of the dephasing source.

Three distinct IFM schemes were investigated. The first system is a concatenation of interferometers based on the integer quantum Hall interferometric architecture proposed in Ref.~[\onlinecite{GiovannettiPRB08}]. The dynamics of electrons traveling along edge channels is exposed to the action of an external fluctuating field. We suggest to steer the propagation of one channel towards the inner part of the Hall bar, where dephasing is minor or absent, and by separating and recombining many times the two channels we reproduce an electronic analogue of the high-efficiency scheme proposed in optics by P. Kwiat {\it et al.} in Ref.~[\onlinecite{KwiatPRL95}]. 
We showed that, for a strong dephasing source, only an asymptotically negligible amount of coherent signal is lost by proper tuning the degree of admixture of the channels at the beam splitters. Moreover, the effect is very robust against small fluctuation about the exact value of the admixture required. Indeed, although the fraction of coherent signal is reduced in magnitude by the averaging process, its qualitative behavior is not affected by it. 

The second system we considered is based on a standard quantum-Hall electronic Mach-Zehnder interferometer and assumes the presence of a quantized electron emitter. A very precisely time-resolved electronic wave packet is sent into a Mach-Zehnder interferometer in which an arm is affected by external classical noise. The packet travels at a precise speed and tests the region affected by noise many times, being split and recombined until it is allowed to escape the interferometer to be collected. The entire sequence 
can be mapped to the concatenation in the space domain that characterizes the noise-sensitive coherent electron channel previously described: the same results and conclusions apply also to this system. The latter has the advantage that it is experimentally much easier to realize, since it is based on a system already available.

The last system we considered is a double-ring structure based on the proposal suggested in Ref.~[\onlinecite{NOI09}]. There, authors considered an Aharonov-Bohm chiral ring in which a localized source of noise affects one arm of the ring and studied the fraction of coherent signal that exits the device. However, such a quantity is not measurable in that setup. We suggest to embed the chiral AB ring in one arm of a larger AB ring and measure the total current flowing through the device as a figure of merit of the coherence of the output signal from the small chiral AB ring. Such a setup has the advantage to overcome the difficulties arising from concatenating many interrogation steps, necessary in order to achieve high efficiency IFM in the noise-sensitive coherent electron channel. It also eliminates the need for very precise time-resolved electronics, on which the second proposal was based.

We point out here that IFM can be designed also for the case of an electron absorber and the same results obtained with the dephasing source are found. The different implementations described here can find useful applications in quantum-coherent electronics and quantum computations, where the coherence of the signals is always threatened by the presence of fluctuating external fields.

This work was supported by funding from the German DFG within SPP 1285 ``Spintronics", from the Swiss SNF via grant no. PP02-106310, and by the Italian MIUR under the FIRB IDEAS project ESQUI. V.~P.~acknowledges CNR-INFM for funding through the SEED Program.

\appendix

\section{Electron-hole switch}
\label{ehswitch}
\begin{figure}[t]
\begin{center}
Ê\includegraphics[width=8cm]{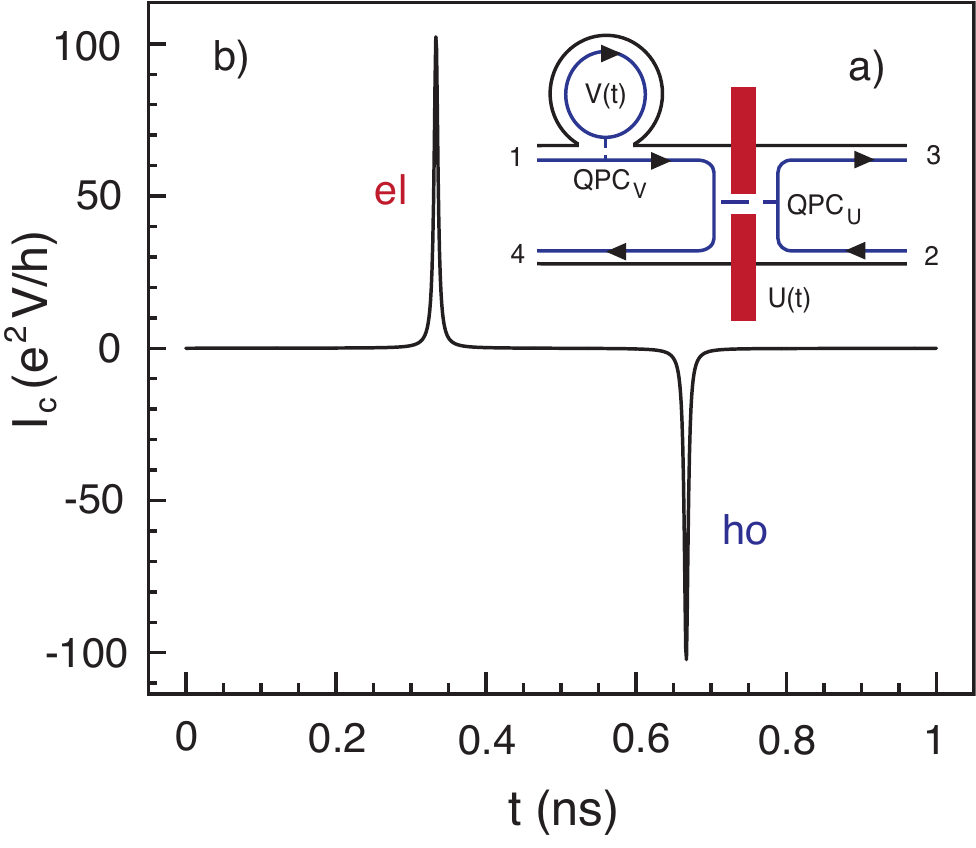}
ÊÊÊ\caption{a) Schematic representation of a time-dependent 
ÊÊÊ		electron-hole switch. The cavity driven by the potential $V(t)$ is 			
		connected via $QPC_V$ to a linear edge and produces a well 				
		separated pair of electron and hole per cycle. The potential $U(t)$ 			
		drives the $QPC_U$ that connects contacts 1 and 2 to contacts 3 and 		
		4 and periodically transmits the electron to contact 3 and reflects 
		the hole to contact 4. b) Time-resolved electron-hole current 				
		produced by the driven cavity in front of $QPC_V$, as given by 
		Eq.~(\ref{Eq:ElHoI}).\label{Fig10}}
\end{center}
\end{figure}

Let us consider the mechanism suggested in Sec.~\ref{Sec4} for injecting and collecting electrons in the MZ interferometer. The system is depicted in Fig.~\ref{Fig10} a) and is composed 
by a cavity formed by a circular edge state that is coupled to an edge channel by a QPC$_V$ of transmission amplitude $\tilde{t}$ and reflection amplitude $\tilde{r}$. 
It was experimentally demonstrated\cite{Gabelli06, Feve07} that such 
a device, if periodically driven by a time-dependent potential $V(t)$, produces 
a periodic current composed by an electron in one half-period and a hole in the other half-period, see Fig.~\ref{Fig10} b). We wish to separate the electron and the hole by transmitting the electron through a barrier towards contact 3, and reflecting the hole into contact 4. A time-dependent QPC$_U$ driven by an external potential $U(t)$ behaves like a beam splitter that mixes the incoming channels, from the contact 1 and 2, into the outgoing channels 3 and 4. If properly driven, it works as a switch that separates electrons and holes generated by the cavity into different edge channels. Following Refs.~[\onlinecite{Moskalets08,Splettstoesser08}] we describe the effect of the time-dependent potential QPC$_U$ by a scattering matrix
\begin{equation}
S_U(t)=\left(\begin{array}{cc}
S_{31}(t) & S_{32}(t)\\
S_{41}(t) & S_{42}(t)
\end{array}\right).
\end{equation}
In the symmetric case one has $S_{31}(t)=S_{42}(t)$ and 
$S_{32}(t)=S_{41}(t)$. From the unitarity of $S_U(t)$ follows that 
\begin{eqnarray}
1&=&\sum_j|S_{jk}(t)|^2,\\
0&=&S_{32}^*(t)S_{31}(t)+S_{42}^*(t)S_{41}(t).
\end{eqnarray}
The dynamics of the cavity can be described by a time-dependent scattering amplitude $S_c(t,E)$, which satisfies $|S_c(t,E)|^2=1$. In the adiabatic regime, keeping all the reservoirs at the same chemical potential $\mu$, the zero-temperature current 
in contacts 3 and 4 can be written as
\begin{equation}
I_j(t)=|S_{j1}(t)|^2I_c(t)+\frac{e}{2\pi i}\sum_{k=1,2}S_{jk}(t)
\frac{\partial}{\partial t}
S_{jk}^*(t),
\end{equation}
with $j=3,4$. Here $I_c(t)$ is the current produced by the cavity, that can be written 
as\cite{Moskalets08,Splettstoesser08}
\begin{equation}\label{Eq:ElHoI}
I_c(t)=\frac{e}{2\pi i}S_c(t,\mu)\frac{\partial}{\partial t}
S_c^*(t,\mu).
\end{equation}
$I_c(t)$ is plotted in Fig.~\ref{Fig10} b) for a harmonic driving 
$V(t)=V_0\cos (\Omega t)$, for the choice $\Omega/2\pi=1~{\rm GHz}$ 
and $|\tilde{t}|^2=0.1$.
By defining $S_{31}(t)=\sqrt{T(t)}$ and $S_{41}(t)=i\sqrt{1-T(t)}$, it follows that $I_3(t)=T(t)I_c(t)$ and $I_4(t)=(1-T(t))I_c(t)$, with $T(t)$ related to the applied external potential $U(t)$. By choosing a proper modulation of $T(t)$, it is possible to separate the electrons from the holes.

\section{Eigenvalue problem}
\label{App1}

Defining
\begin{equation}\label{EqApp-upm}
u_{\pm}=\frac{1}{2\tan(\gamma)}\left(1-{\rm sinc}(\epsilon)\pm
\sqrt{(1+{\rm sinc}(\epsilon))^2-4\frac{{\rm sinc}(\epsilon)}{\cos^2(\gamma)}}\right)
\end{equation}
The matrix $U$ assumes the simple form 
\begin{equation}\label{App1-Eq:U}
U=\left(\begin{array}{cccc}
1 & 0 & 0 & 0\\
0 & 1 & 0 & 0\\
0 & 0 & u_+ & u_-\\
0 & 0 & 1 & 1
\end{array}\right),
\end{equation}
with ${\rm sinc}(\epsilon)=\sin(\pi\epsilon)/\pi\epsilon$, 
that allows for a simple solution of the eigenvalue problem in terms of a Jordan decomposition, ${\cal Q}=U^{-1}{\rm diag}(1,\sin(\pi\epsilon)/\pi\epsilon,\lambda_-,\lambda_+)U$, with
\begin{eqnarray}\label{App1-Eq:lambda}
\lambda_{\pm}&=&\frac{1}{2}\cos(\phi)(1+{\rm sinc}(\epsilon))\nonumber\\
&\pm&\frac{1}{2}\sqrt{\cos^2(\phi)(1+{\rm sinc}(\epsilon))^2-{\rm sinc}^2(\epsilon)}.
\end{eqnarray}

\section{Double ring transmission and reflection amplitudes}
\label{App:TrRef}

In the absence of decoherence, the transmission amplitude for electrons going from the left lead $L$ to the right lead $R$ can be calculated through the following multiple scattering formula
\begin{equation}
\tau=\boldsymbol{\tau}_B(\openone-{\bf \Gamma})^{-1}S'_p\boldsymbol{\tau}_A,
\end{equation}
with ${\bf \Gamma}=S_p'\bar{\boldsymbol{\rho}}_A\bar{S}'_p\boldsymbol{\rho}_B$, and Ê
$S'_p=\left(\begin{array}{cc}
S_p & 0\\
0 & 1\end{array}\right)$.
We define the following transmission matrices in node $A$ and $B$ that take into account the lower arm of the larger ring,
\begin{eqnarray}
{\bf t}'_A&=&\left(\begin{array}{cc}
{\bf t}_A & 0\\
0 & 1\end{array}\right),\qquad
{\bf t}'_B=\left(\begin{array}{cc}
{\bf t}_B & 0\\
0 & 1\end{array}\right),\\
\bar{\bf t}'_A&=&\left(\begin{array}{cc}
\bar{\bf t}_A & 0\\
0 & 1\end{array}\right),\qquad
\bar{\bf t}'_B=\left(\begin{array}{cc}
\bar{\bf t}_B & 0\\
0 & 1\end{array}\right),
\end{eqnarray}
with ${\bf t}'_A$ and $\bar{\bf t}'_B$ of dimension $3\times 2$, and Ê$\bar{\bf t}'_A$ and ${\bf t}'_B$ of dimension $2\times 3$. Analogously we define the reflection matrices 
\begin{eqnarray}
{\bf r}'_A&=&\left(\begin{array}{cc}
r_A & 0\\
0 & 0\end{array}\right),\qquad
\bar{\bf r}'_B=\left(\begin{array}{cc}
\bar{r}_B & 0\\
0 & 0\end{array}\right),\\
\bar{\bf r}'_A&=&\left(\begin{array}{cc}
\bar{\bf r}_A & 0\\
0 & 0\end{array}\right),\qquad
{\bf r}'_B=\left(\begin{array}{cc}
{\bf r}_B & 0\\
0 & 0\end{array}\right),
\end{eqnarray}
with ${\bf r}'_A$ and $\bar{\bf r}'_B$ of dimension $2\times 2$, and Ê$\bar{\bf r}'_A$ and ${\bf r}'_B$ of dimension $3\times 3$.
The effective transmission amplitudes $\boldsymbol{\tau}_A$ and $\boldsymbol{\tau}_B$ are given by the matrices
\begin{eqnarray}
\boldsymbol{\tau}_A&=&{\bf t}'_A
\left(\openone-P\bar{\bf r}_L\bar{P}{\bf r}'_A
\right)^{-1}P{\bf t}_L,\\
\boldsymbol{\tau}_B&=&{\bf t}_L
\left(\openone-P\bar{\bf r}'_B\bar{P}{\bf r}_R
\right)^{-1}P{\bf t}'_B,
\end{eqnarray}
with dimension respectively $3\times 1$ and $1\times 3$. The effective reflection amplitudes $\bar{\boldsymbol{\rho}}_A$ and $\boldsymbol{\rho}_B$ are given by the matrices
\begin{eqnarray}
\bar{\boldsymbol{\rho}}_A&=&\bar{\bf r}'_A+{\bf t}'_AP
\left(\openone-\bar{\bf r}_L\bar{P}{\bf r}'_AP
\right)^{-1}\bar{\bf r}_L\bar{P}\bar{\bf t}'_A,\\
\boldsymbol{\rho}_B&=&{\bf r}'_B+\bar{\bf t}'_B\bar{P}
\left(\openone-{\bf r}_RP\bar{\bf r}'_B\bar{P}
\right)^{-1}{\bf r}_RP{\bf t}'_B.
\end{eqnarray}

\end{document}